\documentclass{article}
\usepackage{spconf,amsmath,graphicx}

\usepackage{hyperref}       
\hypersetup{
    colorlinks,
    linkcolor={red!90!black},
    citecolor={green!90!black},
    urlcolor={black}
}
\usepackage{url}            
\usepackage{booktabs}       
\usepackage{amsfonts}       
\usepackage{xcolor}         
\usepackage{amssymb}
\usepackage{makecell}
\usepackage{multirow}
\usepackage{multicol}

\newcommand{\ie}{\textit{i}.\textit{e}.}

\usepackage{arydshln}

\newcommand\freefootnote[1]{%
  \let\thefootnote\relax%
  \footnotetext{#1}%
  \let\thefootnote\svthefootnote%
}

\usepackage[
backend=biber,
style=ieee,
citestyle=numeric-comp,
maxbibnames=3,
maxcitenames=3,
doi=false,isbn=false,url=false,eprint=false
]{biblatex}
\defbibheading{bibliography}{\begin{center}\textbf{6. REFERENCES}\end{center}}

\title{Text-driven Talking Face Synthesis\\by Reprogramming Audio-driven Models}
\name{Jeongsoo Choi, Minsu Kim, Se Jin Park, Yong Man Ro$^*$\thanks{$^*$Corresponding Author. This work was partially supported by the National Research Foundation of Korea (NRF) grant funded by the Korea government (MSIT) (No. NRF-2022R1A2C2005529) and BK21 FOUR (Connected AI Education \& Research Program for Industry and Society Innovation, KAIST EE, No. 4120200113769).}}
\address{School of Electrical Engineering, KAIST, South Korea}

\begin{document}
\ninept
\maketitle
\begin{abstract}
In this paper, we present a method for reprogramming pre-trained audio-driven talking face synthesis models to operate in a text-driven manner. Consequently, we can easily generate face videos that articulate the provided textual sentences, eliminating the necessity of recording speech for each inference, as required in the audio-driven model. To this end, we propose to embed the input text into the learned audio latent space of the pre-trained audio-driven model, while preserving the face synthesis capability of the original pre-trained model. Specifically, we devise a Text-to-Audio Embedding Module (TAEM) which maps a given text input into the audio latent space by modeling pronunciation and duration characteristics. Furthermore, to consider the speaker characteristics in audio while using text inputs, TAEM is designed to accept a visual speaker embedding. The visual speaker embedding is derived from a single target face image and enables improved mapping of input text to the learned audio latent space by incorporating the speaker characteristics inherent in the audio. The main advantages of the proposed framework are that 1) it can be applied to diverse audio-driven talking face synthesis models and 2) we can generate talking face videos with either text inputs or audio inputs with high flexibility.
\end{abstract}
\section{Introduction}
\label{sec:intro}
With the great development of deep learning, talking face synthesis using deep neural networks \cite{gafni2021dynamic, liang2022expressive, gao2023high} has become more natural and acceptable. The technology is utilized in diverse applications such as virtual avatars \cite{grassal2022neural, wu2023ganhead}, visual dubbing \cite{kim2019neural, xie2021towards}, and teleconferencing \cite{wang2021one}. The mainstream research has been primarily audio-driven \cite{zhou2019talking, chen2019hierarchical, zhou2020makelttalk, prajwal2020lip, thies2020neural, zhou2021pose, guo2021ad, park2022synctalkface, zhua2023audio, shen2023difftalk}, altering the source face video to speak the speech contained in the input audio. In recent years, audio-driven talking face synthesis has achieved high generation quality and accurate lip synchronization between audio and the generated video. However, there is an impediment to fully employing the audio-driven talking face synthesizer in the real world, the difficulty of editability. Since the input of the synthesizer is speech audio, the user should record the speech which is to be said by the avatar, and this could be burdensome to perform for every video to generate.

One simple approach for alleviating this problem can be using Text-to-Speech (TTS) system \cite{ren2021fastspeech, lee2023imaginary}. Since TTS system can generate synthetic speech from a text, it can be utilized to edit the generated results of audio-driven talking face synthesis. However, as it requires two independent model passes (\textit{i.e.}, TTS and audio-driven talking face synthesis), the mismatch between synthetic audio obtained from TTS and real speech audio used to train is inevitable, and this makes the generated results erratic unless performing the fine-tuning of the trained audio-driven talking face synthesis model on the speech audio of TTS.

Another possible direction for bypassing this problem can be generating a talking face video from the text input directly. However, text-video pair datasets are scarce compared to audio-video pair datasets, thus generated videos from the text are less natural compared to audio-driven output. Due to these challenges, text-driven talking face synthesis is not much explored yet compared to audio-driven synthesis. Prior works \cite{chen2020duallip, liu2022parallel, zhang2022text2video} that tried to synthesize the talking face directly from text input are limited to generating video captured in a constrained environment. Thus, they might fail in generating lip movements of arbitrary identities in the wild.

In this paper, we try to reprogram pre-trained audio-driven talking face synthesis models with text inputs, enabling the generation of talking face video in the wild with easy editability using text. To this end, we propose a novel method that converts the input text into audio representations and utilizes the high-quality generation ability of pre-trained audio-driven talking face synthesis models. The proposed model projects text representations into the audio latent space of the pre-trained model, and the projected text representations are directly utilized to synthesize talking face video with the generator of the audio-driven model. Thus, we can edit the lips by simply inserting the sentences without learning the entire text-driven talking face synthesis model from scratch. Moreover, to accurately map the text modality into the learned audio modal space, we propose to inject the identity information from a single face image during the text encoding, which is helpful in modeling the identity characteristics lying in speech audio. Finally, the proposed framework can be applied to various audio-driven talking face synthesis systems \cite{prajwal2020lip, zhou2021pose, shen2023difftalk} which are easily accessible online, without limits on network architectures.

\begin{figure*}[t]
	\centering
	\centerline{\includegraphics[width=\linewidth]{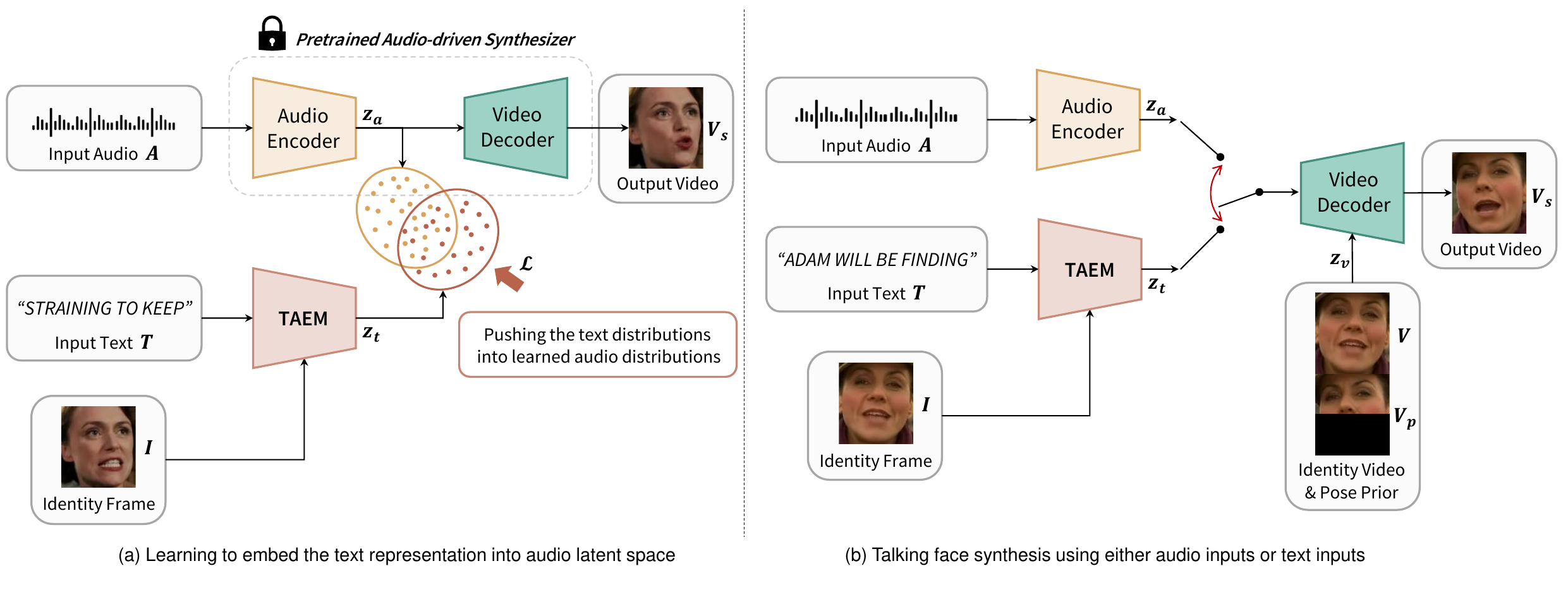}}
	\vspace{-0.5cm}
	\caption{Overview of the proposed text-driven talking face synthesis framework which reprograms the audio-driven models. (a) In the training stage, TAEM learns to embed the text representation into the audio latent space. (b) In the inference stage, we can generate a talking face video by inserting either a text or audio containing desired speech content.}
	\label{fig:1}
 \vspace{-0.4cm}
\end{figure*}

Our contributions can be summarized as follows: we (i) facilitate simple yet effective text-driven talking face synthesis from various pre-trained audio-driven models, enabling the synthesis of face video with either text or audio inputs interchangeably, (ii) propose to use a face image to reflect speaker characteristics for individuals thereby making projection from text to audio more accurate, and (iii) achieve a state-of-the-art performance compared to text-driven talking face synthesis methods and even competitive results to the audio-driven methods, in the wild.

\section{Proposed Method}
Our objective in this paper is allowing to fully utilize the pre-trained audio-driven talking face synthesis model by introducing editability to the model with text inputs. The text is closely related to speech audio semantically, so humans can easily associate the relationship between text and audio. We explore whether it is possible to replace the input from audio to text without affecting the original performance of the audio-driven pipeline. The main hypothesis of our method is that text can be appropriately represented in the audio latent space which is trained with the talking face synthesis objective. Therefore, the goal of our learning problem is projecting the text representations into the audio latent space thereby talking face synthesizer can generate video with the nearly same performance as the pre-trained audio-driven synthesizer. Thus, after training, it is able to utilize both modalities, text and audio inputs interchangeably to generate talking face videos. The overall pipeline of the proposed method is illustrated in Fig. \ref{fig:1}.

\vspace{-0.2cm}
\subsection{Baseline Audio-driven Model}
Although the proposed method can be applied to various audio-driven talking face synthesis models, we describe our method based on applying to Wav2Lip \cite{prajwal2020lip}, one of the most popular audio-driven talking face synthesis models.
Let $V$ be an input video for providing identity information and $A$ be an input audio containing target content. The input video and audio are embedded into feature vectors $z_v=f_v(V,V_p)$ and $z_a=f_a(A)$, respectively, by using fully-convolutional video encoder $f_v$ and audio encoder $f_a$, where $V_p$ is upper face video for providing the target pose. The generator of Wav2Lip takes visual features $z_v$ in the intermediate layers and $z_a$ as inputs to synthesize the output talking face video $V_s$.

\vspace{-0.2cm}
\subsection{Text-to-Audio Embedding Module (TAEM)}
\begin{figure}[t]
	\begin{minipage}[b]{1.0\linewidth}
		\centering
		\centerline{\includegraphics[width=\linewidth]{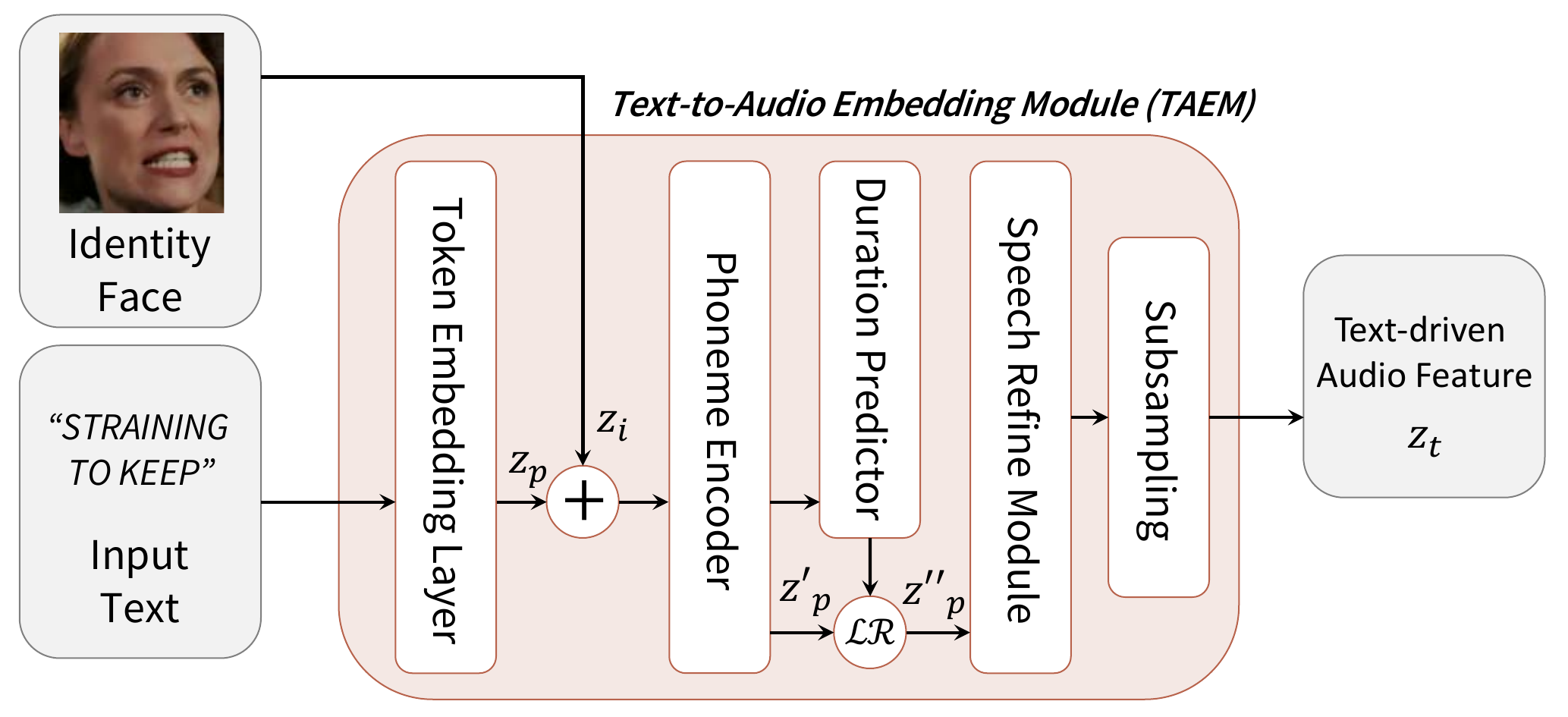}}
	\end{minipage}
	\vspace{-0.5cm}
	\caption{The detailed architecture of the proposed TAEM.}
    \vspace{-0.4cm}
	\label{fig:2}
\end{figure}

When the audio-driven talking face model is given, our objective is to substitute the audio features $z_a$ with text representations $z_t=f_t(T)$, where the text representations are embedded by a proposed Text-to-Audio Embedding Module (TAEM), $f_t$, using input text $T$. The proposed TAEM is guided to encode the text feature $z_t$ to be aligned with the latent space of learned audio representations $z_a$. The overall architecture of TAEM $f_t$ is shown in Fig. \ref{fig:2}.

\vspace{-0.2cm}
\subsubsection{Face Embedding and Phoneme Encoder}
Firstly, TAEM takes a phoneme sequence converted from raw text input. The phoneme sequence is mapped into dense vectors $z_p$ with a dimension of $\mathbb{R}^{l_t\times512}$ through a token embedding layer, where $l_t$ is the length of the phoneme sequence. Since $z_p$ does not hold the identity information that exists in the audio, such as personal voice characteristics, we introduce a speaker embedding extracted from the input face to accurately model the acoustic features from the text. As personal voice is correlated with the face as explored in \cite{chung20c_interspeech,hong2022visagesyntalk,lee2023imaginary}, the face feature can be utilized as speaker embedding. We utilize a pre-trained face recognition model ArcFace \cite{deng2019arcface} to extract the visual speaker embedding $z_i$ from a face image $I\in V$. Then, the speaker embedding is added to the phoneme feature of each timestep to incorporate personal characteristics. A phoneme encoder $g$ converts the speaker-embedded phoneme features into acoustic features $z_p'$. For the phoneme encoder, we employ a Transformer-based network \cite{vaswani2017attention} whose feed-forward network contains a 1D convolution layer, following \cite{ren2021fastspeech}. Phoneme encoder can model the context by considering neighbor phonemes through self-attention.

\vspace{-0.2cm}
\subsubsection{Duration Predictor and Length Regulator}
The obtained acoustic features $z_p'$ do not hold accurate time information which should be modeled to generate accurate talking face video. To this end, we introduce a duration predictor which predicts the time length of each phoneme in the input text. The inputs of duration predictor $\mathcal{DP}$ are acoustic features and it generates duration $d$ for each feature vector similar to \cite{ren2021fastspeech, liu2022parallel} as $d=\mathcal{DP}(z_p')$.
The duration $d \in \mathbb{R}^{l_t}$ is used for extending feature vectors in the temporal dimension. To this end, a length regulator $\mathcal{LR}$ is used which repeats acoustic features by the predicted duration $d$ so that each phoneme can be temporally aligned with audio. For example, when $d=\{3,1,2\}$ and $z_p=\{z_{p1},z_{p2},z_{p3}\}$, then the aligned acoustic features can be represented as $\mathcal{LR}(z_p, d)=\{z_{p1},z_{p1},z_{p1},z_{p2},z_{p3},z_{p3}\}$. We make the total length of acoustic features equal to $l_a$, the number of mel-spectrogram frames, so that the text-to-audio mapping can be constructed. Obtaining the aligned acoustic features $z_p''$ can be written as follows, $z_p''=\mathcal{LR}(z_p', d)$.

\vspace{-0.2cm}
\subsubsection{Speech Refine Module}
Since the aligned acoustic features $z_p''$ have repetition in value, the generated frames from $z_p''$ can have temporally sticking lip shapes, and discontinuities between the frames. Thus, a Transformer based speech refine module is introduced. The speech refine module $h$ converts $z_p''$ to the smooth audio speech vectors with a dimension of $\mathbb{R}^{l_a\times512}$ by considering context information in $z_p''$. As we use a single audio feature vector when generating one frame of synthesized video whose number of frames is $l_v$, we need to temporally subsample $z_p''\in\mathbb{R}^{l_a\times512}$ into $z_t\in\mathbb{R}^{l_v\times512}$. To this end, a subsampling network adjusts the length of the vector sequence and produces text-driven audio features $z_t$. Briefly, the production of text-driven audio features can be represented as $z_t = \text{subsample}(h(z_p''))$.

\vspace{-0.1cm}
\subsection{Objective Functions}
The proposed TAEM is basically optimized with mean squared error (MSE) loss between the audio feature and the mapped text-driven audio feature. Then, TAEM can map the text inputs into the learned audio latent space of the pre-trained audio-driven talking face synthesis model. The objective function can be written as follows,
\begin{equation}
\setlength{\abovedisplayskip}{3pt}
\setlength{\belowdisplayskip}{3pt}
    \mathcal{L}_{dis}=\frac{1}{l_v}\sum_{i}^{l_v}||z_{a,i}-z_{t,i}||^{2}_2.
\end{equation}
We examine different types of loss functions and show that simple loss (\ie, MSE) is sufficient for generating precise video.

Along with $\mathcal{L}_{dis}$, a duration loss is used to train the duration predictor $\mathcal{DP}$. The error between the predicted duration and the ground-truth duration is minimized $\mathcal{L}_{dur}=||d-\hat{d}||^{2}_2$.
We stop gradient from input feature vectors for disconnecting duration modeling from text-to-audio learning, thereby $\mathcal{DP}$ is optimized explicitly.

\vspace{-0.2cm}
\section{Experimental Setup}
\subsection{Dataset}
\textbf{GRID} \cite{cooke2006audio} is a phrase-level dataset of 33 speakers for 1000 utterances each.  Each phrase contains 6 words from a fixed dictionary. Each video lasts 3 seconds with frame rate 25fps. We use the same seen speaker setting of \cite{liu2022parallel}. 

\noindent \textbf{TCD-TIMIT} \cite{harte2015tcd} is a sentence-level dataset. Different from GRID, each sentence is phonetically rich and does not have a fixed vocabulary or a fixed grammar. Each video lasts 2.5$ \sim $8.1 seconds and the frame rate is 29.97 fps. The total duration is about 7.5 hours. We use the `volunteers' subset and `straight' camera angle following previous work \cite{liu2022parallel}. 

\noindent \textbf{LRS2} \cite{afouras2018deep} is a sentence-level dataset of 140,000 utterances from TV shows from BBC. It is considered a challenging set because it contains thousands of speakers without speaker labels and large variations in head poses. Please note that we first utilize the LRS2 dataset which was not exploited due to its difficulties in a text-driven talking face generation task.

\begin{figure}[t!]
	\begin{minipage}[b]{1.0\linewidth}
		\centering
		\centerline{\includegraphics[width=\linewidth]{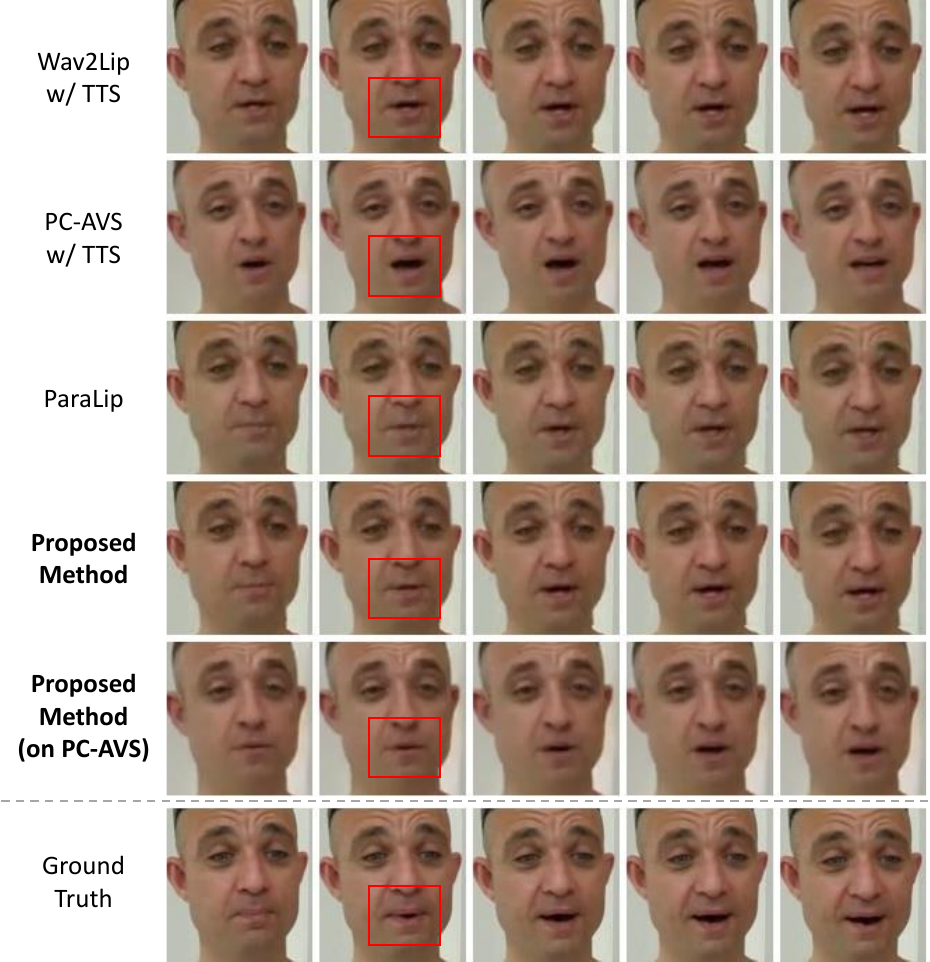}}
	\end{minipage}
	\vspace{-0.6cm}
	\caption{Qualitative results comparison on LRS2 dataset.}
	\vspace{-0.4cm}
	\label{fig:3}
\end{figure}

\subsection{Baseline Methods}
We compare our method with three talking face synthesis methods. \textbf{ParaLip} \cite{liu2022parallel} is a text-driven approach of text-to-lip generation through a parallel decoding model. But as ParaLip generates only the lip region, we additionally implement a video decoder on our own so that it generates the entire face video. We take two audio-driven methods, \textbf{Wav2Lip} \cite{prajwal2020lip} and \textbf{PC-AVS} \cite{zhou2021pose}. Since they are purely audio-based, we also explore their text editability by using TTS to provide the input speech. We use FastSpeech2 \cite{ren2021fastspeech} model pre-trained on the LJSpeech dataset for the TTS. Since it is a serial form of the TTS and the audio-driven talking face synthesizer, we named the methods using TTS as cascaded models. In order to compare the generated videos with ground-truth videos, we provide ground-truth duration to synthesize audio. For all three methods, we use open-source codes to train the models on the target datasets.

\subsection{Evaluation Metrics}
We use PSNR, SSIM, LMD, LSE-D, and LSE-C to evaluate the method. PSNR and SSIM are visual quality metrics while LMD, LSE-D, and LSE-C are lip-sync quality metrics. LMD measures the distance between lip landmarks of the ground truth faces and those of the generated faces. LSE-C and LSE-D are the confidence score and the distance score of the features from a pre-trained SyncNet \cite{chung2017out} model proposed by \cite{prajwal2020lip}. For fair comparisons, we first crop the face using dlib \cite{dlib09} and then evaluate the same regions of the face.

\begin{table*}[t]
  \renewcommand{\arraystretch}{1.4}
  \renewcommand{\tabcolsep}{1.0mm}
  \centering
  \vspace{-0.2cm}
  \caption{Quantitative results on GRID, TCD-TIMIT, and LRS2 dataset. $\uparrow$: the higher is the better; $\downarrow$: the lower is the better.}
  \vspace{0.1cm}
  \resizebox{\linewidth}{!}{
  \begin{tabular}{cc ccccc ccccc ccccc}
    \Xhline{3\arrayrulewidth}
    \multirow{2}{*}{\makecell{\textbf{Input}\\ \textbf{Modality}}} & \multirow{2}{*}{\textbf{Method}} & \multicolumn{5}{c}{\textbf{GRID}} & \multicolumn{5}{c}{\textbf{TCD-TIMIT}} & \multicolumn{5}{c}{\textbf{LRS2}} \\ \cmidrule(l{2pt}r{2pt}){3-7} \cmidrule(l{2pt}r{2pt}){8-12} \cmidrule(l{2pt}r{2pt}){13-17}
    & & \textbf{SSIM}$\uparrow$ & \textbf{PSNR}$\uparrow$ & \textbf{LMD}$\downarrow$ & \textbf{LSE-D}$\downarrow$ & \textbf{LSE-C}$\uparrow$ &
    \textbf{SSIM}$\uparrow$ & \textbf{PSNR}$\uparrow$ & \textbf{LMD}$\downarrow$ & \textbf{LSE-D}$\downarrow$ & \textbf{LSE-C}$\uparrow$ &
    \textbf{SSIM}$\uparrow$ & \textbf{PSNR}$\uparrow$ & \textbf{LMD}$\downarrow$ & \textbf{LSE-D}$\downarrow$ & \textbf{LSE-C}$\uparrow$\\ \cmidrule(l{2pt}r{2pt}){1-2} \cmidrule(l{2pt}r{2pt}){3-7} \cmidrule(l{2pt}r{2pt}){8-12} \cmidrule(l{2pt}r{2pt}){13-17}

    \multirow{2}{*}{\textbf{Audio}} & Wav2Lip \cite{prajwal2020lip} & 0.920 & 33.748 & 0.872 & 7.009 & 6.324
            & 0.859 & 32.506 & 0.875 & 6.855 & 5.564
            & 0.868 & 32.044 & 1.020 & 5.014 & 9.556 \\
    & PC-AVS \cite{zhou2021pose} & 0.775 & 30.027 & 1.335 & 6.944 & 6.577
            & 0.715 & 30.182 & 1.381 & 6.330 & 6.486
            & 0.688 & 29.736 & 1.590 & 6.560 & 7.770 \\ \cmidrule(l{2pt}r{2pt}){1-2} \cmidrule(l{2pt}r{2pt}){3-7} \cmidrule(l{2pt}r{2pt}){8-12} \cmidrule(l{2pt}r{2pt}){13-17}
    \multirow{4}{*}{\textbf{Text}} & Wav2Lip w/ TTS & 0.913 & 33.537 & 1.080 & 8.088 & 5.244
                   & 0.845 & 32.216 & 1.223 & 8.859 & 4.546
                   & 0.863 & 31.986 & 1.122 & 7.026 & 7.430 \\
    & PC-AVS w/ TTS & 0.768 & 30.008 & 1.577 & 8.750 & 4.992
                  & 0.701 & 30.099 & 1.797 & 9.029 & 4.647
                  & 0.650 & 29.554 & 1.532 & 8.238 & 6.016 \\
    & ParaLip$^\dagger$ \cite{liu2022parallel} & 0.919 & 33.343 & \textbf{0.881} & 7.305 & 5.952
                      & \textbf{0.859} & 32.477 & \textbf{0.906} & 7.082 & 5.727
                      & 0.865 & \textbf{32.046} & 1.070 & 6.828 & 7.378 \\ \cdashline{2-17}
    & \textbf{Proposed Method}
    & \textbf{0.920} & \textbf{33.734} & 0.885 & \textbf{7.014} & \textbf{6.306} 
    & \textbf{0.859} & \textbf{32.485} & 0.909 & \textbf{6.985} & \textbf{5.886}
    & \textbf{0.867} & 32.011 & \textbf{1.055} & \textbf{5.826} & \textbf{8.763} \\
    \Xhline{3\arrayrulewidth}
  \end{tabular}
  }
  \label{table:1}
  \vspace{-0.4cm}
\end{table*}

\subsection{Implementation Details}
The input text is tokenized at the phoneme level. We use Montreal Forced Aligner (MFA) \cite{McAuliffe2017} to find the time-alignment map between the phoneme and the speech. We extract the ground-truth duration of each phoneme from the time-alignment map. For preprocessing video, we obtain a face ROI utilizing dlib \cite{dlib09}, and each frame is cropped into the face region. The cropped frame is then resized into the size of 96 $\times$ 96. Audio is converted into Mel-spectrogram using a hop size of 160 and a window size of 800, constructing 100 fps. $g$ and $h$ in our proposed TAEM contain 4 blocks of Feed Forward Transformer borrowed from \cite{ren2021fastspeech}. Sinusoidal positional encoding is used for both of them. The hidden vector size is set to 512. The random face image (\ie, $I$) of the input video is used to extract speaker embedding. The dimension of the feature vector is 512 so that it can be element-wisely added to the phoneme vectors. Duration predictor contains two 1D convolutional layers and one fully-connected layer to generate one number for each input vector. Subsampling layer contains two 1D convolutional layers and stride 2 is utilized. We use AdamW optimizer with $\beta_1=0.9$, $\beta_2=0.98$, $\epsilon=10^{-8}$ when training our proposed TAEM. The batch size of 64 and learning rate of 0.0001 are utilized. The model is implemented on PyTorch and we use one 48GB RTX A6000 GPU. We use the audio latent feature from the audio encoder of Wav2Lip as the target when training TAEM. The video encoder and decoder for generating output video are attached to the TAEM during inference.

\vspace{-0.3cm}
\section{Experimental Results}
\vspace{-0.1cm}
\subsection{Generation Quality Comparison}
To verify the effectiveness of the proposed method, we compare the generation quality of different methods. The comparison results on GRID, TCD-TIMIT, and LRS2 are shown in Table \ref{table:1}. Compared to the audio-driven talking face synthesis models (\ie, Wav2Lip and PC-AVS), the proposed method achieves comparable results even with text inputs. This shows that the proposed method has no loss in the generation quality when the model is reprogrammed to use text inputs. Moreover, the proposed method outperforms the cascaded methods (\ie, Wav2Lip w/ TTS and PC-AVS w/ TTS) on both visual quality and lip sync quality, showing the superiority of the proposed method over the cascaded methods. Finally, by comparing with a text-driven talking face synthesis model, ParaLip, the proposed method shows better performance than the previous method overall. Please note that different from ParaLip, the proposed method can be applied to any audio-driven model. Therefore, the proposed text-driven framework can be steadily improved along with the advance of the audio-driven model. Moreover, as we reprogram the audio-driven model instead of building a new model, the proposed method converges about 5 times faster than ParaLip (5k steps vs. 25k steps). Qualitative results are shown in Fig.~\ref{fig:3}. The man is pronouncing `but' where the jaw slightly drops. The results show that the proposed method best captures the gradual opening of the mouth which is enclosed by the red box. More qualitative results can be found in our demo page\footnote{\label{note1}\textbf{Demo page}: generated samples can be found on \href{https://bit.ly/45R9LGo}{bit.ly/45R9LGo}}.

\freefootnote{\hspace{-0.13cm}$^\dagger$ParaLip is re-implemented to generate face, not lip region.}

\begin{table}[t!]
	\renewcommand{\arraystretch}{1.4}
	\renewcommand{\tabcolsep}{2.5mm}
 \centering
 \vspace{-0.2cm}
 \caption{Ablation study about the loss functions on LRS2 dataset. $\uparrow$: the higher is the better; $\downarrow$: the lower the better.}
 \vspace{0.1cm}
\resizebox{0.9999\linewidth}{!}{
\begin{tabular}{cccccc}
\Xhline{3\arrayrulewidth}
\textbf{Loss} & \textbf{SSIM} $\uparrow$ & \textbf{PSNR} $\uparrow$ & \textbf{LMD} $\downarrow$ & \textbf{LSE-D} $\downarrow$ & \textbf{LSE-C} $\uparrow$\\ \hline
Contrastive & 0.801 & 31.077 & 1.204 & 7.068 & 7.648 \\
MSE & \textbf{0.867} & 32.011 & 1.055 & \textbf{5.826} & 8.763 \\
MSE+Contrastive & 0.866 & 32.005 & 1.059 & 5.978 & \textbf{8.805} \\
MSE+Task & 0.866 & \textbf{32.018} & \textbf{1.050} & 5.841 & 8.718 \\
\Xhline{3\arrayrulewidth}
\end{tabular}}
\label{table:2}
\vspace{-0.4cm}
\end{table}

\subsection{Ablation Study}
We conduct an ablation study to investigate the effectiveness of different loss functions for $\mathcal{L}_{dis}$. Table \ref{table:2} shows the ablation results according to different loss functions. `MSE' means MSE loss used as $\mathcal{L}_{dis}$ in this paper. `Contrastive' means contrastive loss that makes positive pairs be pulled closer while negative pairs are pushed farther from each other in the latent space \cite{oord2018representation}. `MSE+Contrastive' is the combination of the two losses. Finally, we examine the effect of the task loss that was originally utilized to pre-train the audio-driven talking face synthesis model. Task loss contains L1 loss between synthesized video and ground-truth video, and sync loss proposed in \cite{prajwal2020lip}. We observe severe artifacts in the upper face when using `Contrastive' loss only. Moreover, using more loss functions (\ie, MSE+Contrastive and MSE+Task) does not lead to much better results. Since the results are not that different from those of `MSE' only, we employ MSE loss for $\mathcal{L}_{dis}$.

\subsection{Application to Another Method}
The main advantage of the proposed method is that it can be applied to diverse audio-driven talking face synthesis models. In order to confirm this, we engage our method with PC-AVS \cite{zhou2021pose}. PC-AVS concatenates identity, pose, and speech content features and uses these features in StyleGAN2 \cite{karras2020analyzing} generator to synthesize the video. We use a publicly available PC-AVS model pre-trained on VoxCeleb2 \cite{chung18b_interspeech} dataset in order to adopt the proposed method. The generated results can be found on the 5th row in Fig. \ref{fig:3} and the demo page\textsuperscript{\ref{note1}}. As the proposed method learns the mapping of text-to-audio, it does not depend on the specific model architectures and can be applied to reprogram the audio-driven model into text-driven.

\vspace{-0.1cm}
\section{Conclusion}
\vspace{-0.1cm}
In this paper, we propose Text-to-Audio Embedding Module (TAEM) to reprogram a pre-trained audio-driven talking face synthesis model to be operated with text inputs. With the proposed method, we can easily manipulate the pre-trained audio-driven synthesis by just typing a sentence. We show that the proposed TAEM can maintain the high video quality of the pre-trained model and achieve better performance than the previous cascaded and text-driven method.


\printbibliography[heading=bibliography]

\end{document}